%% file: main.tex
\documentclass[letterpaper]{article} %
\usepackage{aaai25}  %
\usepackage{times}  %
\usepackage{helvet}  %
\usepackage{courier}  %
\usepackage[hyphens]{url}  %
\usepackage{graphicx} %
\usepackage{natbib}  %
\usepackage{caption} %
\usepackage{inconsolata}

\input{macros}
\usepackage{mathtools}
\usepackage{multirow}
\usepackage{array,booktabs}

\usepackage{adjustbox}
\usepackage{algorithm}
\usepackage{algpseudocodex}

\usepackage{newfloat}
\usepackage{listings}
\DeclareCaptionStyle{ruled}{labelfont=normalfont,labelsep=colon,strut=off} %
\floatstyle{ruled}
\newfloat{listing}{tb}{lst}{}
\floatname{listing}{Listing}
\title{Priority Matters: Optimising Kubernetes Clusters Usage\\ with Constraint-Based Pod Packing}

\author {
    Henrik Daniel Christensen\textsuperscript{\rm 1},
    Saverio Giallorenzo\textsuperscript{\rm 2},
    Jacopo Mauro\textsuperscript{\rm 1}
}
\affiliations {
    \textsuperscript{\rm 1}University of Southern Denmark\\
    \textsuperscript{\rm 2}University of Bologna\\
    hench13@student.sdu.dk , saverio.giallorenzo@unibo.it, mauro@imada.sdu.dk
}

\begin{document}

\maketitle

\input{sections/00_abstract}

\input{sections/01_introduction}

\input{sections/02_preliminaries}

\input{sections/03_implementation}

\input{sections/04_evaluation}

\input{sections/05_related_work}

\input{sections/06_conclusions}

\section{Acknowledgements}

We thank Simone Canova, Alessandro Neri, and Matteo Trentin, who contributed to
preliminary groundwork for the techniques presented herein. Their initial
explorations and suggestions helped the foundation behind this research.\todo{remove for IJCAI submission, include in arxiv version}

\bibliography{biblio}

\end{document}

%% file: macros.tex
\usepackage[
  disable,     %
  textwidth=\marginparwidth,
  colorinlistoftodos,
  prependcaption,
  linecolor=green,backgroundcolor=green!25,bordercolor=green
  ]{todonotes}
\makeatletter
\if@todonotes@disabled
\else %
\newlength{\increase}
\paperwidth=\dimexpr \paperwidth + \increase\relax
\oddsidemargin=\dimexpr\oddsidemargin + .5\increase\relax
\evensidemargin=\dimexpr\evensidemargin + .6\increase\relax
\marginparwidth=\dimexpr \marginparwidth + .4\increase\relax
\let\tmptitle\title
\renewcommand{\title}[1]{\tmptitle{#1{\tiny\normalfont \makebox[10em]{\colorbox{red!20}{disable todos}}}}}
\fi
\makeatother

\usepackage{amsmath}
\usepackage[capitalize,noabbrev]{cleveref}
\usepackage{nameref}

%% file: sections/00_abstract.tex
\begin{abstract}
Distributed applications employ Kubernetes for scalable, fault-tolerant
deployments over computer clusters, where application components run in groups
of containers called pods. The scheduler, at the heart of Kubernetes'
architecture, determines the placement of pods given their priority and resource
requirements on cluster nodes. To quickly allocate pods, the scheduler uses
lightweight heuristics that can lead to suboptimal placements and resource
fragmentation, preventing allocations of otherwise deployable pods on the
available nodes.

We propose the usage of constraint programming to find the optimal allocation of
pods satisfying all their priorities and resource requests. Implementation-wise,
our solution comes as a plug-in to the default scheduler that operates as a
fallback mechanism when some pods cannot be allocated. Using the OR-Tools
constraint solver, our experiments on small-to-mid-sized clusters indicate that,
within a 1-second scheduling window, our approach places more higher-priority
pods than the default scheduler (possibly demonstrating allocation optimality)
in over 44\% of realisable allocation scenarios where the default scheduler
fails, while certifying that the default scheduler's placement is already
optimal in over 19\% of scenarios. With a 10-second window, our approach
improves placements in over 73\% and still certifies that the default
scheduler's placement is already optimal in over 19\% of scenarios.

\end{abstract}

%% file: sections/01_introduction.tex
\section{Introduction}

The state-of-the-art paradigms for implementing distributed systems are
microservices~\cite{DGLMMMS17} and serverless~\cite{JSSTKePSCKYGPSP19}. These
paradigms decompose a distributed application into components that one can
replicate independently of each other, according to their workload, avoiding
wasting resources on overprovisioned components. Practically, developers and
platforms use containers~\cite{M14} to package these components and simplify
their deployment/replication and container orchestrators~\cite{JBBFMMP19} to
automate deployment, scaling, and management of containers on clusters of
computation nodes. Nowadays, the main container orchestrator is
Kubernetes~\cite{L18}. In Kubernetes, the \emph{pod} represents the smallest
deployable unit, enclosing a group of containers that share an execution context
(e.g., networking, storage).

\begin{figure}[t]
 \vspace{-1em}
 \includegraphics[width=\columnwidth]{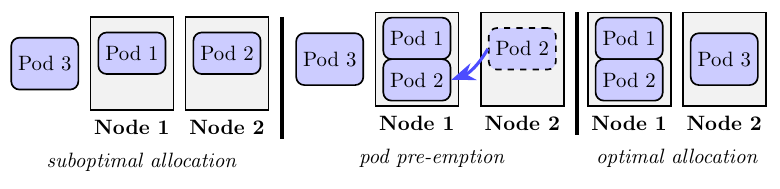}
 \vspace{-1em}
 \caption{\label{fig:intro}Example of Kubernetes scheduler suboptimal pod
 allocation problem -- Node and Pod sizes represent their memory
 capacity/occupancy. Left: the heuristics place Pod 1 and Pod 2 on separate
 nodes, leaving insufficient memory on either node for Pod 3, despite the
 cluster having enough total resources. Centre: reallocation of the first
 two pods onto one node enables the (Right:) optimal placement of all
 pods.\vspace{-1em}}
\end{figure}

The scheduler, at the heart of Kubernetes' architecture, determines pods'
placement according to their requirements and node resources.
The scheduler determines allocations through heuristics that prioritise speed
and scalability but neglect optimal resource usage, possibly leading to system
degradation (some pods are not allocated) or overprovisioning (the inclusion of
new cluster nodes to host the pods not allocated) and resource waste (money,
energy, etc.)~\cite{clickhouse_saving_nodate}. Recent
studies~\cite{CastAI2025KubernetesCostBenchmark} have shown that a staggering
99.94\% of Kubernetes clusters are overprovisioned, where the average gap
between provisioned and requested resources is 40\% for CPU and 57\% for memory. 

As an illustration of this phenomenon, shown in \cref{fig:intro}, let us consider a scenario with a
2-node cluster, each node with 4 GB of memory, where we want to deploy 3 pods in
sequence, all with the same priority but requiring respectively 2, 2, and 3 GB
of memory. When scheduling a pod, the scheduler heuristics use a
\texttt{LeastAllocated} policy to rank nodes. Thus, it likely allocates the
first two pods on one node each. When the scheduler receives the request for the
last pod, it cannot find a node with 3 GB of memory available, leaving the
allocation request pending, possibly requiring the addition of a new node to
deploy the pod. The allocation (left of \cref{fig:intro}) is \emph{suboptimal}
because the nodes have enough resources to host all pods. If we could reallocate
one of two pods onto one node (centre of \cref{fig:intro}), we would obtain the
optimal allocation (right of \cref{fig:intro}).

Pod priorities further complicate the problem of finding the optimal schedule in
Kubernetes. The scheduler must allocate high-priority pods preferentially, even
if it means pre-empting queued lower-priority pods. Hence, the dynamic nature of
workloads, which requires the frequent creation/deletion of pods with varying
priorities, further complicates the task of making allocation decisions that
optimise overall resource usage and performance. In the worst case, the default
scheduler's policy can lead to pod
starvation.\footnote{\url{https://github.com/kubernetes/kubernetes/issues/86373}}

\paragraph{Our contribution}

We tackle the shortcomings of the Kubernetes scheduler heuristics by identifying
failed deployment attempts that could fit within the cluster resources and
devising an optimal resource allocation that ensures efficient resource usage
and wards off avoidable operational costs. In this instance, we focus our
attention on the representative case of small and medium-size clusters, which
are renownedly prone to overprovisioning~\cite{clickhouse_saving_nodate-1} and
where even saving single-node resources can significantly impact overall
resource availability and costs, especially when leveraging the cloud
infrastructure.

Our proposal is a conservative one and works by complementing the heuristics of
the Kubernetes scheduler. We let the default heuristics handle allocations as
long as they can satisfy pod requirements. If scheduling requests become pending
(signalling requirement unsatisfiability) but they can fit within the available
resources, then we use a (complete) solver to possibly find a solution (within a
fixed timeframe) that satisfies all the priority and resource constraints.

To find a global optimum, we solve a series of optimisation problems on the
possible pod allocations, given the pods' priority. For every priority level, we
guarantee the deployment of the maximum number of pods with that priority and
minimise the total number of moved pods, to curtail overall system disruptions.
The allocation problem is notoriously NP-hard~\cite{DBLP:books/fm/GareyJ79},
which implies that finding an optimal solution may take a long time. Since we
aim to have a scheduling decision within a useful timeframe (e.g., 10 sec.\@),
we set a timeout for each priority level to provide either a proven optimum
or, if the solver timed-out, the best feasible allocation found until
the timeout.

We evaluate feasibility and limitations of this approach using the
state-of-the-art CP-SAT solver from OR-Tools~\cite{PF24}. We simulate scheduling
decision requests for small-to-medium-sized clusters with up to 32 nodes and
varying degrees of pod requests, resource demands, and priorities.
Our findings demonstrate that our approach scales to medium-sized clusters and
can usually find a better, if not optimal, solution in less than 1 seconds for
cluster of up to 8 nodes, or in 10--20 seconds for clusters of up to 32 nodes.
In particular, with 1 second timeouts (resp. 10 seconds), our approach finds
better or optimal solutions in 44\% (resp. 73\%) of scenarios where the default
scheduler can not allocate all the pods. Moreover, our approach proves that the
solution of the default scheduler is already optimal and can not be improved in
19\% (resp. 19\%) cases.

\emph{Structure of the paper}
We introduce Kubernetes and its scheduler in \nameref{sec:preliminaries}. In
\nameref{sec:implementation} we detail our optimisation approach, its
realisation, and integration as a Kubernetes plugin. In \nameref{sec:evaluation}
we present the experiments conducted to evaluate our approach. We conclude by
discussing \nameref{sec:related} and future directions.

%% file: sections/02_preliminaries.tex
\section{Preliminaries}
\label{sec:preliminaries}

\subsubsection{Kubernetes}

Kubernetes is the leading container orchestration platform for running cloud
application workloads. In Kubernetes, scaling cloud architecture components,
like microservices or serverless functions, typically entails running on a
cluster of nodes one or more pod instances -- each containing one or more
software components -- to obtain a desired number of available component replicas.

Kubernetes allows developers to assign a priority to every pod, useful to
control pod pre-emption, i.e., the removal of pods from a node to make space for
higher-priority ones. In such scenarios, Kubernetes sends a signal to a
pre-empted pod to gracefully terminate -- the default ``grace period'' is 30
seconds -- after which the pod is (forcefully) removed.

To control the placement of pods, Kubernetes uses labels and selectors. Users
can assign labels to nodes and pods and express (anti-)affinity constraints to
influence the scheduling. For example, if a pod benefits from being scheduled on
a node with an SSD, we can specify it to be affine with the label
\texttt{ssd-disk} that also marks the SSD-equipped nodes.
A Kubernetes cluster includes a \emph{control plane} and one or more nodes. The
control plane exposes the API and interfaces to define, deploy, and manage the
life cycle of containers. The nodes are the machines that host the containers.
Each node hosts two components, \texttt{kubelet} -- an agent responsible for
actively running the pods and checking their status -- and \texttt{kube-proxy}
-- a proxy that manages networking rules on nodes and supports network communication.

In production, the control plane usually runs across multiple machines and a
cluster runs multiple nodes, providing fault-tolerance and high
availability~\cite{noauthor_kubernetes_nodate-1}. The control plane takes global
decisions about the cluster, as well as detecting and handling cluster events,
like errors. One of control plane's central components is the scheduler.

\subsubsection*{Scheduler}

The scheduler decides where to place pods based on a set of rules and the
available resources of the cluster. The default scheduler is called
\texttt{kube-scheduler}.

Nodes that meet the requirements for a pod are called ``feasible''.
\texttt{kube-scheduler}'s decision process includes two successive phases:
filtering and scoring. During filtering, \texttt{kube-scheduler} selects the
feasible nodes, checking both resource and (anti-)affinity requirements. At the
scoring, the scheduler finds the best node for placing the pod.

The scheduling framework consists of a set of ``plugin'' APIs directly compiled
into the scheduler. These APIs allow one to implement most scheduling features
as plugins while keeping the scheduling ``core'' lightweight and
maintainable~\cite{noauthor_scheduling_nodate}. Two cycles determine the
scheduling workflow, the \emph{scheduling} and the \emph{binding}. The
scheduling cycle selects a node for placing the pod. The binding cycle links the
pod to the selected node and prevents future scheduling cycles from allocating
other pods on the portion of the node reserved to the linked pod.

\begin{figure}[t]
\centering
\includegraphics[width=\columnwidth]{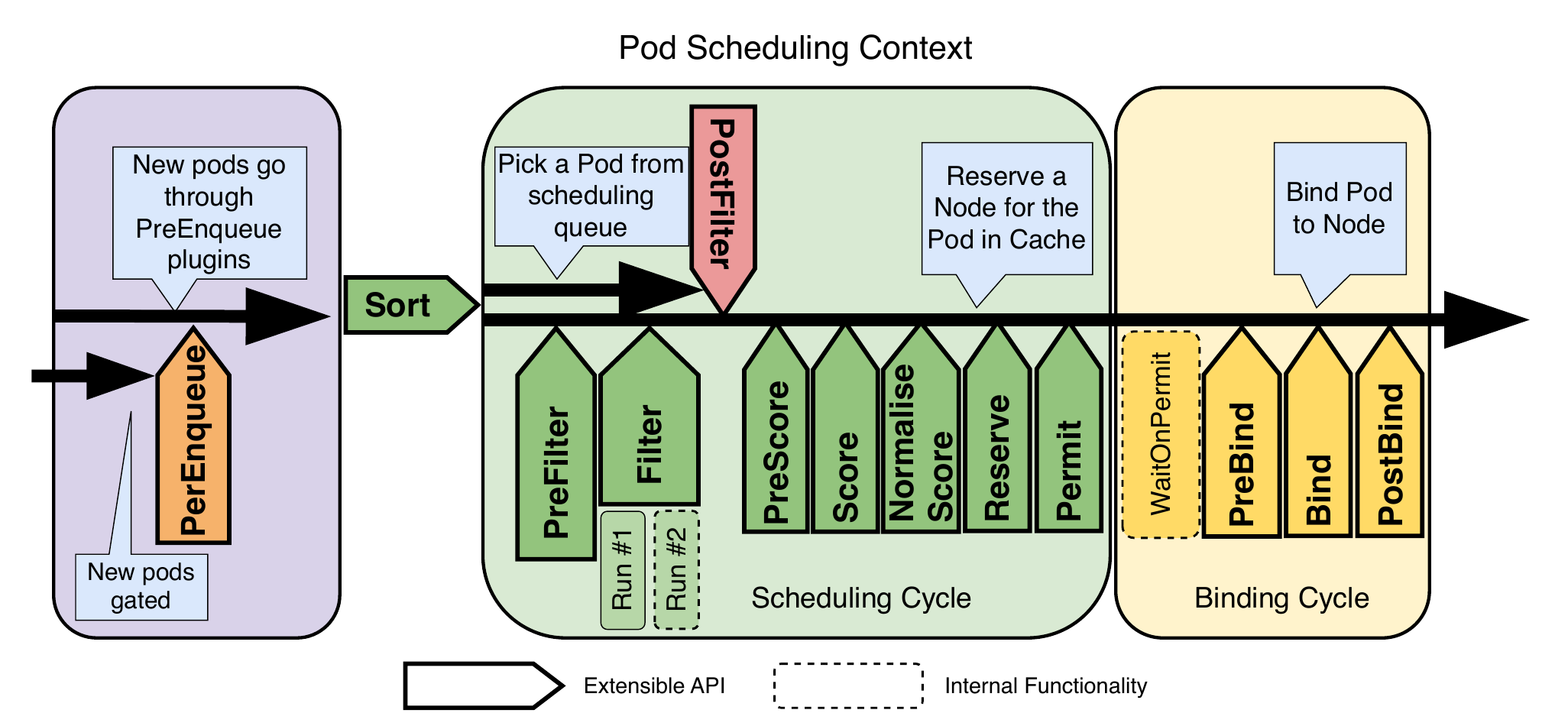}
\caption{Scheduling Framework Extensions Points.}
\label{fig:scheduling-framework-extensions}
\end{figure}

\cref{fig:scheduling-framework-extensions}\footnote{Adapted from https://kubernetes.io/docs/concepts/scheduling-eviction/scheduling-framework/.}, illustrates the scheduling framework,
where pentagons represent extension points and dashed rectangles indicate
internal functionalities (unmodifiable); left to right:

\begin{itemize}
\setlength\itemsep{-.2em}
\item 
\texttt{PreEnqueue} performs checks on the pod before it goes into the
ready-for-scheduling queue.
If one of these checks fails, the pod is not marked \texttt{Schedulable} (nor
marked \texttt{Unschedulable}), and it goes in a special queue.

\item 
\texttt{QueueSort} sorts pods in the scheduling queue. Only one of such plugins can be active at a time.

\item 
\texttt{PreFilter} pre-processes information about the pod or checks conditions
about the cluster. If one of these plugins returns an error, the scheduling cycle fails.

\item 
\texttt{Filter} performs the filtering phase and prunes the nodes where the pod cannot run.

\item
\texttt{PostFilter} plugins activate only if the \texttt{Filter} plugins mark
all nodes of the cluster as infeasible. This extension point is used mainly for pre-emption purposes.

\item \texttt{Score} implements the scoring phase.

\item \texttt{NormalizeScore} adjusts the scores produced by the \texttt{Score}
plugins before completing the final ranking.

\item \texttt{Reserve} activates before binding a pod to its designed node and
takes care of concurrency issues that can emerge from parallel scheduling sessions.
If the \texttt{Reserve} plugin, or a following one, fails, the resources reserved
by this plugin are released by the \texttt{Unreserve} plugin.

\item \texttt{Permit} delay the binding of a pod to its selected node.

\item \texttt{PreBind} prepares the cluster (and the selected node) before the
pod-node binding. If a \texttt{PreBind} plugin fails, the pod is rejected and
placed in the scheduling queue.

\item \texttt{Bind} plugins actively bind a pod to a node and they can choose
whether to handle the pod or not. If a plugin does not bind the pod, another
\texttt{Bind} plugin runs until either one successfully executes or the binding
cycle gets cancelled, and the pod goes back into the scheduling queue.

\item \texttt{PostBind} allow for final checks or modifications before the pod
is bound to the selected node.

\end{itemize}

%% file: sections/03_implementation.tex
\section{Implementation}
\label{sec:implementation}

We describe the formalisation of the optimisation problem we designed and how we
integrate the formalisation into a proof-of-concept Kubernetes scheduler plugin.

\subsubsection{Optimisation problem}

We solve the optimisation problem incrementally in a loop that iterates over all
priorities. In every priority tier, a solver runs twice: first to maximise the
number of placed pods, and second to minimise the number of allocated pods to
move between nodes.

Notation: we use $n \in N$ to denote nodes. We define with $n.\text{ram}$ and
$n.\text{cpu}$ the capacities of $n$. For each pod $p \in P$, let $p.\text{ram}$
and $p.\text{cpu}$ be its resource requests, $p.\text{priority} \in [0,
pr_{\max}]$ its priority level (with lower values denoting higher priority), and
$p.\text{where}$ the index of the node where the pod is currently placed
(\(p.\text{where}=0\) if the pod is not scheduled). The binary variables $x_{i,
j} \in \{0, 1\}$ indicate whether pod $p_i$ is assigned to node $n_j$.

We report in \cref{alg:opt} the pseudocode of the optimisation loop,
where a model $m$ collects constraints and supports a solver that maximises a given metric.

\algrenewcommand{\algorithmiccomment}[1]{%
  \hfill \textit{\color{gray}\(\triangleright\) #1}}

\begin{algorithm}
\caption{\label{alg:opt}Optimisation Algorithm.}
\begin{algorithmic}[1]
\State $m \gets \text{Model}()$
\For{$pr \in \text{range}(p_{\max}+1)$}
    \State $m.\text{add\_constraint}(\text{bin\_packing\_constraints}(pr))$
    \State\Comment{(1) maximise number of placed pods}
    \State $\text{metric} \gets \sum_{p_i \in P: p_i.\text{priority} \leq pr} \sum_{n_j \in N} x_{i,j}$
    \State $\text{sol} \gets m.\text{max}(\text{metric}, \text{get\_timeout}())$
    \If{$\text{sol}.\text{status} = \text{OPTIMAL}$}
        \State $m.\text{add\_constraint}(\text{metric} = \text{sol}(\text{metric}))$
    \Else
        \State $m.\text{add\_constraint}(\text{metric} \geq \text{sol}(\text{metric}))$
    \EndIf
    \State\Comment{(2) minimise number of pod moves}
    \State $\text{metric} \gets$
    \State \phantom{A}$\sum_{\substack{p_i \in P: p_i.\text{priority} \leq pr \\ \wedge\ p_{i}.\text{ where} \neq 0}} \left( \sum_{n_j \in N} x_{i,j} + 2 x_{i, p_i.\text{where}} \right)$
    \State $\text{sol} \gets m.\text{max}(\text{metric}, \text{get\_timeout}())$
    \If{$\text{sol}.\text{status} = \text{OPTIMAL}$}
        \State $m.\text{add\_constraint}(\text{metric} = \text{sol}(\text{metric}))$
    \Else
        \State $m.\text{add\_constraint}(\text{metric} \leq \text{sol}(\text{metric}))$
    \EndIf
\EndFor
\State \Return $\text{sol}$
\end{algorithmic}
\end{algorithm}

First, we initialise the model without constraints (Line 1). Then, for each
priority level \(pr\) from the highest (0) to the lowest (\(p_{max}\)) we add
(multi-dimensional) bin-packing constraints (Line 3) to restrict solutions only
to valid placements, and use these to (1) maximise the pods allocated up to
priority \(pr\) (Lines 5--10), and (2) to minimise the removal of already
deployed pods (Lines 12--18).

In \cref{alg:opt}, operation ``bin\_packing\_constraints(pr)'', inspired by
\citet{DBLP:conf/cp/Shaw04}, indicates the addition of constraints that ensure
that the resource capacities of nodes are not exceeded when placing pods with
priority up to \(pr\). Formally, we add the following constraints.
\begin{align}
    \label{shawn1}\sum_{p_i\in P:\, p_i.\mathrm{priority}\,\le\,pr} x_{i, j} \cdot p_i.\text{ram} \leq n_j.\text{ram} \quad \forall j \in B \\
    \label{shawn2}\sum_{p_i\in P:\, p_i.\mathrm{priority}\,\le\,pr} x_{i, j} \cdot p_i.\text{cpu} \leq n_j.\text{cpu} \quad \forall j \in B \\
    \label{shawn3}\sum_{j \in N} x_{i, j} \leq 1 \quad \forall p_i\in P:\, p_i.\mathrm{priority}\,\le\,pr
\end{align}

Constraints (\ref{shawn1}) and (\ref{shawn2}) impose that the sum of the
requested resources by the pod with priority up to \(pr\) are within the
respective node's RAM and CPU capacities, while (\ref{shawn3}) ensures the
assignment of a pod with priority up to \(pr\) to at most one
node.\footnote{Unlike~\citet{DBLP:conf/cp/Shaw04}, we omit the constraint that
the sum of the bin loads is equal to the sum of the item size, which would
prevent us from computing the optimal schedule if the resources were not enough
to host all the pods -- our core problem is not a bin-packing one and closer to
a multi-knapsack one. We have also considered adding symmetry-breaking
constraints~\cite{DBLP:conf/cp/Shaw04}, but we found that they did not improve
the solving time in our experiments.}

At Line 6 of \cref{alg:opt}, we maximise the number of pods within the current
priority by maximising of the metric
\[\sum_{p_i \in P:\, p_i.\mathrm{priority}\le pr}\sum_{n_j \in N} x_{i,j}.\]

If the solver finds an \texttt{OPTIMAL} solution, we update the model by adding
a constraint that imposes to find future solutions with exactly the same amount
of pods up to priority \(pr\) (Line 8). If the solver did not terminate within
the allocated time, thus not proving that the solution is optimal but providing
only a \texttt{FEASIBLE} solution, we add a constraint that imposes to find
solutions with at least the same amount of pods up to priority \(pr\) (Line 10).

In the second stage, we minimise the disruptions caused by avoidable pod
evictions. To do so, we minimise the evictions of allocated pods up to priority
\(pr\) by asking the solver, at Line 14, to maximise the following metric:
\[
\sum_{p_i \in P:\, p_i.\mathrm{priority}\le pr \land p_i.\mathrm{where}\ne 0} \left(\sum_{n_j \in N} x_{i,j} + 2x_{i,p_i.\mathrm{where}}\right)
\]
where we give a weight of 2 to the pods that remain in place (i.e., assigned to
the same node they are already allocated onto) and a weight of 1 to the pods
to move to another node.

Note that, since bin-packing is an NP-hard problem, both problems solved at
every iteration of the loop are NP-hard, and the solver could take a long time
to find an optimum. Since we define our algorithm to find a placement for pods,
we model it to provide a solution (possibly not optimal) in a reasonable amount
of time. For this reason, we design it to run with a fixed wall-clock timeout
$T_{\mathrm{total}}$, which limits the total solving time across all priority
tiers. To avoid that a single priority tier exhausts the entire time budget,
thus compromising the possibility to find a solution involving also lower
priorities pods, we dedicate a fraction $\alpha \in [0,1]$ of the total time to
each priority tier, leaving the remaining $(1-\alpha) T_{\mathrm{total}}$ time
for the opportunistic termination of the solver execution. Moreover, for every
priority, we split in half the reserved time between the two solving phases
(maximisation of pods and minimisation of evictions). In particular, in
\cref{alg:opt}, we use function $\texttt{get\_timeout}$ at every solver
invocation (Line 6 and 14) to compute the available time budget. Formally,
assuming that \texttt{unused} is the amount of unreserved time yet to consume,
$\alpha \in [0,1]$
\[\texttt{get\_timeout}() = \alpha T_{\mathrm{total}}/({p_{max}}+1) +
\texttt{unused} \]

\paragraph{Solver}

We implement our approach by using OR-Tools~\cite{PF24} -- an open-source suite
developed by Google for optimisation tasks using, among other approaches, a
lazy-constraint approach. Notably, OR-Tools is the winner of the latest MiniZinc
Challenge~\cite{minizinc_challenge}, a competition evaluating the performance of
constraint programming solvers across diverse benchmarks. In particular, we use
the OR-Tools' CP-SAT Python API, which runs several complementary search
strategies in parallel. Since, unlike popular SMT solvers (e.g., Z3), CP-SAT
does not support incremental push/pop of constraints, we re-solve the model
after each place/move step. To try to speed up the search, we use CP-SAT hints
to warm-start the next solve by providing the current assignment.

\paragraph{Kubernetes Plugin}

We recall that our goal is to conservatively enhance Kubernetes' scheduler by
enabling optimal scheduling decisions while retaining the efficiency of the
default scheduling process. Unlike prior
approaches~\cite{lebesbye_boreas_2021,luca_sage_2023}, which strive for
optimality at every decision, our design invokes the solver only
periodically or when needed (e.g., via an HTTP API), allowing the
default scheduler to efficiently handle most allocations.

Notably, achieving global optimality requires pre-empting pods across nodes. At
the moment Kubernetes pre-emption operates within a single node and API support
to cross-node pod pre-emption is under consideration for
development~\cite{noauthor_scheduler-pluginspkgcrossnodepreemption_nodate,
crossnodepreemption_no_implementation}. To implement of our
proposal, we develop a custom scheduler plugin capable of evicting and
re-scheduling pods according to the solver's optimised placement, including
cross-node preemption.

Given the current Kubernetes APIs, we implement the atomic binding
and pre-emption of the pods that our optimiser suggests through the
scheduling and eviction in separate scheduling events.
We implement our plugin across five scheduler framework extension points:
\texttt{PreEnqueue}, \texttt{PreFilter}, \texttt{PostFilter}, \texttt{Reserve/Unreserve}, and
\texttt{PostBind} (cf.~\nameref{sec:preliminaries}). Default preemption is
disabled to ensure that all eviction and relocation decisions are controlled
exclusively by our optimisation logic.

When the optimisation algorithm is invoked, it considers all cluster pods, both
scheduled and pending. If the optimiser proposes relocating certain pods, we
evict and re-schedule them in separate scheduling events, according to the
computed plan. During solver execution, new pods arriving in the scheduling
queue are temporarily paused. We record these pods in an internal list, which is
used to re-queue them once the solver execution completes. The faster the
solver, the smaller is the scheduling latency for new pods.

The plugin manages the allocation of pods affected by the optimiser's plan at
multiple extension points. First, at the \texttt{PreEnqueue} and
\texttt{PreFilter} points, it assigns the affected pods to their target nodes,
allowing the default scheduler to bind them accordingly.
In the \texttt{PostFilter} we mark the pods that cannot be scheduled by the
default scheduler. Since pod names change upon rescheduling, and we cannot rely
on them for correct placement, at the
\texttt{Reserve/Unreserve} point, the plugin explicitly reserves resources for
the affected pods. Finally, at the \texttt{PostBind} point the plugin monitors
progress and marks the plan as completed once all intended allocations are
realised.

We have written our plugin in Go, which is the language used for the development
of Kubernetes, and one can deploy it by extending Kubernetes' default scheduler
container. The optimisation algorithm calls a Python script, which invokes
OR-Tools, deployed within the scheduler's container.
The default scheduler, extended with our plugin, can be deployed as a container
image within the cluster, replacing the original scheduler pod instance.

%% file: sections/04_evaluation.tex
\section{Evaluation}
\label{sec:evaluation}

We move on to the empirical evaluation of our proposal. To the best of our
knowledge, there is no standard benchmark for Kubernetes' scheduler, nor
publicly available datasets of real-world Kubernetes scheduling decision
requests. Hence, to evaluate the performance of our optimisation algorithm, we
generate random traces of pod requests, simulate initial allocations, and record
the performance of our algorithm. To simulate the scheduling process, we use
Kubernetes WithOut Kubelet (KWOK)~\cite{KWOK} -- an open-source simulator that
faithfully mirrors the behaviour of the standard Kubernetes scheduler. KWOK does
not manage actual node resources, and it simulates node capacities and
pod resource requests, enabling large-scale and reproducible scheduling
experiments with minimal overhead.

Specifically, we generate a set of pod requests with configurable \emph{a)}
number of nodes in the cluster, \emph{b)} average number of pods per node,
\emph{c)} workload ratio between the total amount of resources in the cluster
and the ones needed by the pods, and \emph{d)} maximal amount of pods' priorities.

We create the pods with random values of CPU and RAM in the interval $[100,
1000]$. The total sum of these resource demands determines the node capacities
together with the workload ratio. 
To reflect typical cloud deployments,
all nodes are assumed to have identical resource capacities.
Finally, we generate random ReplicaSets requests, i.e., a request to deploy a given amount of replicas of a pod. Each request requires the deployment of a random number in $[1,4]$ of pods. The pod's priorities are generated randomly.

We generate different datasets. Each dataset consists of 100 instances with
various parameters combinations. We considered clusters with 4, 8, 16 or 32
nodes, with 4 and 8 average number of pods per node, and priority tiers in
${1,2,4}$. As target usage, we considered 90\%, 95\%, 100\%, and 105\% of
a cluster resources.
We test the optimisation algorithm with different $T_{\mathrm{total}}$ timeouts:
1, 10, and 20 seconds.

The lower target usage levels (90\% and 95\%) present many instances that the
default scheduler could solve trivially (thus not triggering our plugin). Thus,
to focus on more challenging cases, we discard the instances where KWOK
successfully places all pods, selecting the first 100 instances it fails to do
so. Because the default scheduler is non-deterministic, for reproducibility,
when generating the dataset instances (and only there) we force KWOK to behave
deterministically by i) introducing a lightweight \emph{Score} plugin to order
nodes by their lexicographic name, ii) set \texttt{parallelism=1} to eliminate
concurrency-induced non-determinism, and disable the \emph{DefaultPreemption}
plugin to prevent evicting lower-priority pods during scheduling.

After obtaining the 100 instances for each combination of parameters
configuration, we evaluate our scheduling approach by running the default
scheduler (as-is) in KWOK and then our optimisation algorithm, if the default
scheduler failed to place all pods. We record the placements of pods and
whether the optimiser found an optimal solution or achieved a better allocation
than the KWOK baseline (i.e., higher number of higher-priority pods).

We run all experiments on virtual machines with 8 vCPUs and 48 GB of RAM (Intel
Xeon Gold 6130), Ubuntu Linux, and OR-Tools version 9.14.6206. The source code
of the scheduler with our new plugin, including the optimisation algorithm, is
available at \url{https://github.com/henrikdchristensen/scheduler-plugins} --
the repository includes a thorough README under the directory at
\texttt{pkg/mypriorityoptimizer} and the scripts to generate benchmarks, run
tests, and reproduce the plots.\todo{for submission change this to supplementtal
material}

\subsubsection{Results}

\begin{figure*}[t]
    \centering
    \includegraphics[width=.85\textwidth,trim={0.25cm 0.30cm 0.24cm 0.23cm},clip]{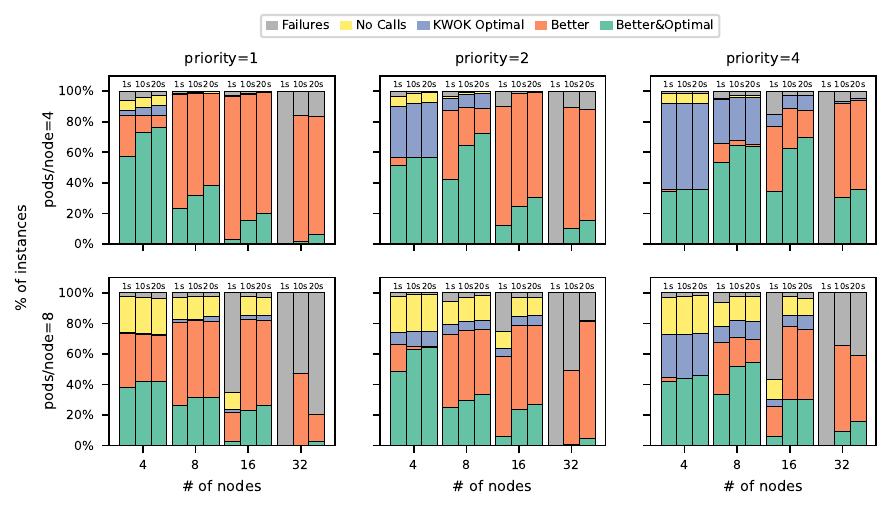}
    \caption{Distribution of solved instances by cluster size (x-axis). We
    collate the results by priority (columns) and pods-per-node (rows),
    aggregating usage levels. For each cluster size, grouped
    bars correspond to increasing solver timeouts (1, 10, 20\,s).}
    \label{fig:2d-stacked-grid-prio-by-ppn}
\end{figure*}

We visualise the summary of the results obtained with our optimisation algorithm
in \Cref{fig:2d-stacked-grid-prio-by-ppn}. In the figure, the x-axis shows the
number of nodes in the cluster, while the y-axis reports the percentage of
instances of a given category (explained below). For every cluster size (4, 8,
16, and 32), we have three grouped stack bars reporting the different results
using incremental solver timeouts (1, 10, and 20\,s). In
\cref{fig:2d-stacked-grid-prio-by-ppn}, we compose the results to show the data
related to a given priority tier (1, 2, 4) and average pods-per-node ratio (4
and 8), aggregating across target usage levels.

The instance categories in the stack bars are:
\begin{itemize}
\setlength\itemsep{-.2em}
\item green (\texttt{Better\&Optimal}), the optimiser found an optimal solution
that was better than the one provided by the default scheduler; 
\item orange (\texttt{Better}), the solver found a better solution, but it could
not terminate before the timeout, thus without proving the solution's
optimality;
\item blue (\texttt{KWOK Optimal}), the default scheduler produced a solution
that the solver proved to be optimal (i.e., no additional pod could be added);
\item yellow (\texttt{No Calls}), the default scheduler allocated all pods and
the solver was not invoked -- due to the non-deterministic nature of the default scheduler;
\item grey (\texttt{Failures}), the solver failed to produce a solution within
the given time limit.
\end{itemize}

As expected, shown in \cref{fig:2d-stacked-grid-prio-by-ppn}, increasing the
timeout generally allows the optimiser to find more optimal solutions. For
example, with 4 pods per node and two priority tiers, the share of optimal
outcomes (green) visibly increases between 1~s and 10~s, with little gain
reaching 20~s. We infer that most instances can be solved within a few seconds,
and longer timeouts mainly help the more complex cases.

The optimisation problem becomes increasingly difficult as the cluster size
grows. When moving from 8 to 32 nodes, the proportion of failed instances rises,
showing that the solver often times-out and find no solution. This effect
reflects the exponential search space increase, as the solver must explore a
much larger number of configurations. In particular, for 32-node clusters, even
a 20~s timeout slightly increases the chances to reach feasibility or
optimality.

The number of pods per node also affects solver performance. With fewer pods per
node, there are fewer possible placements, which makes the problem simpler. The
effect of the number of priorities is also notable. As the number of priority
tiers increases, the optimiser tends to find a higher proportion of optimal
solutions (blue and green bars combined). This result likely comes from the fact
that priorities introduce a clearer ordering between pods, guiding the solver's
search toward better solutions.

Summarising the results from \cref{fig:2d-stacked-grid-prio-by-ppn}, longer
timeouts improve solution quality but with diminishing returns beyond 10~s,
larger clusters and higher pod densities increase solver difficulty, while
additional priorities tend to guide the optimiser toward more optimal outcomes.

\begin{figure}[t]
\centering
\includegraphics[width=\linewidth,trim={0.74cm 0.78cm 0.13cm 0.23cm},clip]{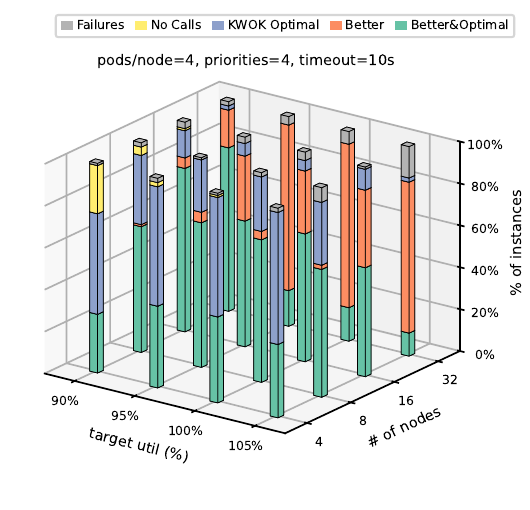}
\caption{Distribution of solved instances for different target usage levels,
with 4 pods per node, 4 priorities, and a solver timeout of
10\,s.}
\label{fig:3d-plots}
\end{figure}

\Cref{fig:3d-plots} shows the target usage levels affect the optimisation
performance as 3D bar plots of the percentage of solved instances, with 4 pods
per node, 4 priorities, and a solver timeout of 10\,s (other combinations show
similar trends).

Overall, the usage level has only a moderate effect on solver performance. While
higher usage levels (100--105\%) slightly increases the number of failed or
non-optimal outcomes, the overall trend is relatively stable across the levels.
At lower usage levels (90--95\%), the default scheduler frequently manages to
schedule all pods (yellow), which limits the number of cases where optimisation
is needed, suggesting that its benefits become more relevant as cluster
resources approach full capacity.

\begin{table}[t]
\centering
\scriptsize
\caption{Solver performance metrics with 4 and 8 pods per node at different target usage levels. The results are based on 4 priorities, 10s solver timeout. The reason that the solver duration exceeds a little the timeout is that we limit the solver itself, and the time here is the total duration including extraction of the solution and I/O, which may slightly be above the solver timeout.}
\label{tab:results-4_8ppn}
\setlength{\tabcolsep}{3pt}
\renewcommand{\arraystretch}{0.88}
\setlength{\aboverulesep}{0pt}
\setlength{\belowrulesep}{0pt}
\begin{tabular}{@{}c l *{8}{c}@{}}
\toprule
\multirow{2}{*}{\textbf{util}} & \multirow{2}{*}{\textbf{metric}} &
\multicolumn{4}{c}{\textbf{ppn = 4}} & \multicolumn{4}{c}{\textbf{ppn = 8}}\\
\cmidrule(lr){3-6}\cmidrule(lr){7-10}
 &  & \textbf{4} & \textbf{8} & \textbf{16} & \textbf{32} & \textbf{4} & \textbf{8} & \textbf{16} & \textbf{32}\\
\midrule
\multirow{3}{*}{90\%}
& solver\,duration\,(s) & 0.9 & 1.2 & 2.4 & 5.9 & 1.0 & 1.1 & 1.8 & \multicolumn{1}{c}{—} \\
& $\Delta$\,cpu\,util\,(\%) & 3.3 & 3.5 & 3.4 & 4.0 & 3.9 & 2.6 & 1.6 & \multicolumn{1}{c}{—} \\
& $\Delta$\,mem\,util\,(\%) & 3.1 & 3.7 & 3.1 & 3.9 & 3.4 & 2.6 & 1.6 & \multicolumn{1}{c}{—} \\
\midrule
\multirow{2}{*}{95\%}
& solver\,duration\,(s) & 0.9 & 1.7 & 5.0 & 9.8 & 1.1 & 1.5 & 3.9 & 10.1 \\
& $\Delta$\,cpu\,util\,(\%) & 2.4 & 3.3 & 3.6 & 4.3 & 2.6 & 2.9 & 2.5 & 0.2 \\
& $\Delta$\,mem\,util\,(\%) & 2.4 & 3.3 & 3.7 & 4.3 & 2.6 & 2.9 & 2.5 & 0.2 \\
\midrule
\multirow{2}{*}{100\%}
& solver\,duration\,(s) & 0.9 & 1.7 & 5.2 & 10.1 & 1.0 & 5.0 & 10.3 & 10.9 \\
& $\Delta$\,cpu\,util\,(\%) & 2.5 & 2.8 & 3.2 & 3.2 & 1.8 & 3.3 & 3.4 & 2.5 \\
& $\Delta$\,mem\,util\,(\%) & 2.1 & 2.5 & 3.0 & 3.3 & 2.1 & 3.1 & 3.1 & 2.4 \\
\midrule
\multirow{2}{*}{105\%}
& solver\,duration\,(s) & 0.9 & 1.6 & 6.0 & 10.2 & 1.2 & 6.2 & 10.3 & 11.0 \\
& $\Delta$\,cpu\,util\,(\%) & 2.1 & 2.6 & 3.2 & 1.9 & 2.5 & 2.3 & 2.6 & 0.6 \\
& $\Delta$\,mem\,util\,(\%) & 1.8 & 2.1 & 2.9 & 2.1 & 2.4 & 2.8 & 2.5 & 0.5 \\
\bottomrule
\end{tabular}
\end{table}

\Cref{tab:results-4_8ppn} presents the average solver duration and the combined
improvement in CPU and memory usage ($\Delta$\,cpu/mem\,util\,(\%)) relative to
the default scheduler. We report the results for configurations with 4 and 8
pods per node, using 4 priority tiers, and a solver timeout of 10\,s.

As expected, the solver duration generally increases with the number of nodes,
reflecting the growing complexity of larger problems. For small cluster sizes
(4--8 nodes), the solver completes within 1--2\,s, while for 16 nodes, solving
time rises to 5--10\,s, hitting the limit of 10\,s at 32 nodes.

The improvement in CPU and memory usage remains almost consistently around
2--4\% across usage targets. This result shows that the optimiser not only
schedules more high-priority pods, but it also achieves denser and more
efficient resource allocations than the default scheduler. Even when the usage
level is already high, the optimiser finds better placements that slightly
increase the cluster capacity. At 100--105\% usage levels, improvements diminish
for the largest clusters, as most pods are already placed and few optimisation
opportunities remain. We also observe that higher numbers of pods per node and
larger cluster sizes tend to slightly reduce usage gains, as the increased
problem complexity limits the solver's ability to find optimal solutions within
the timeout. Finally, as expected, the higher number of priority tiers increases
the solver duration, since we iterate over each priority tier.

%% file: sections/05_related_work.tex
\section{Related Work}
\label{sec:related}

We position our work among the proposal for the optimisation of
Kubernetes' scheduler, referring the interested reader to
\citet{DBLP:journals/eor/DelormeIM16} and
\citet{DBLP:journals/cor/CacchianiILM22a} for a survey on exact algorithms for
the bin-packing and knapsack problems proposed in the last fifty years.

\citet{wei-guo_research_2018} and \citet{carrion_kubernetes_2023} surveyed
Kubernetes' scheduler optimisations with meta-heuristics based on evolutionary
and swarm-intelligence algorithms and particle-swarm optimisation. In general,
these approaches are slower than the default scheduler.

Other works use machine learning techniques
to improve scheduling decisions, predicting the scheduling tasks or the effects
of the allocation of pods on nodes. For example, DL2~\cite{peng_dl2_2021}
uses the data of the default scheduler to train a new scheduler based on a deep
neural network and relies on reinforcement learning to improve it. After an
initial training time, the DL2 scheduler replaces the default scheduler. The
authors of DL2 claim improvements over the default scheduler up to $44.1 \%$ in
terms of average job completion time. However, a key challenge with DL2 and
similar approaches arises when there are changes in workload patterns or cluster
structure. The performance of machine learning can degrade significantly under
such conditions, leading to suboptimal scheduling decisions.

Another direction regards the usage of mathematical modelling to compute the
optimal scheduler decision. Attempts in this area include
SAGE~\cite{luca_sage_2023} and BOREAS~\cite{lebesbye_boreas_2021}.
SAGE is a utility that, given a set of pods and deployment requirements,
determines what set of VMs is optimal to deploy the pods, creates a Kubernetes
cluster on those VMs, and schedules the pods on the nodes. BOREAS is a scheduler
that collects pod requests and uses a configurator optimiser to allocate the
pods on the remaining available nodes of the cluster. 
Both approaches do not consider priorities and pre-emption and rely on solving a
bin-packing problem hinged on the assumption that node resources are enough to
accommodate all the pods.

\citet{santos_towards_2019} present a solution that relies on runtime
information on the cluster and implement an extension of the default scheduling
mechanism for Kubernetes, enabling it to make resource provisioning decisions
based on the current status of the network infrastructure. Similarly,
\citet{nguyen_elasticfog_2020} collect information on the distribution of
network traffic and allocate resources proportionally to it.

\citet{kaur_keids_2020} and \citet{james_low_2019} optimise the scheduling
capabilities of Kubernetes based on power consumption and usage of green energy.
These are interesting extensions that, differently from our proposal, consider
alternative policies to the maximisation of allocated pods.

%% file: sections/06_conclusions.tex
\section{Conclusions}
\label{sec:conclusions}

We present an improvement to Kubernetes' scheduler that makes optimal scheduling
decisions when pod requests cannot be fulfilled and become pending. We show
that, by using off-the-shelf optimisation solvers, it is possible to improve the
default scheduler allocation for Kubernetes clusters of small and medium size
(up to 32 nodes) and with a varying number of pods, resource demands, and
priority levels.

We deem our work a first step for future research in understanding the
algorithmic trade-offs of performing multi-node pre-emption in Kubernetes.
One promising future direction is to explore the usage of other solvers (e.g.,
MIP solvers like Gurobi~\cite{gurobi}, CP solvers like Gecode~\cite{gecode} or
bin-packing solvers of the BPPLib library~\cite{bpp_lib}), possibly using them
in a portfolio approach \cite{DBLP:journals/aim/Kotthoff14} to improve the
chances of finding better solutions. In particular, we envision exploring the
possibility of combining heuristic approaches
\cite{DBLP:journals/jois/MunienE21} meant to find good solutions quickly with
complete solvers that can prove the solution optimal, if more time is available.

Another evolution regards supporting Kubernetes' scheduler rules and cluster
zones (i.e., requirements like node/pod (anti-)affinity that constraint pods
deployment). This work would require adding constraints to further control the
allocation of pods, possibly making the problem easier to solve thanks to a reduction of the search space.